\documentstyle[amssymb]{article}

\input{tcilatex}

\begin{document}

\author{Bertrand Le Roy}
\title{${\Bbb Z}_{n}^{3}$-graded Poincar\'{e} superalgebras}
\date{september 12th, 1997}
\maketitle

\begin{abstract}
Using colored superanalysis and $\varepsilon $-Lie superalgebras, we build
the minimal Poincar\'{e} superalgebra in the case of the ${\Bbb Z}_{n}^{3}$
grading. We then build a representation of this algebra, and the
corresponding Poincar\'{e} supergroup.
\end{abstract}

\newpage

\section{Introduction}

This work is based on generalized Grassmann algebras, which are graded by an
arbitrary abelian group and obey generalized commutation relations. These
relations are based on a commutation factor $\varepsilon $ that is a
function of the degrees of the elements it applies to: 
\[
ab=\varepsilon (d_{a},d_{b})ba
\]
where $d_{a}$ and $d_{b}$ are the degrees of $a$ and $b$, which are elements
of the grading group. This commutation factor obeys several very restricting
rules.

It is well known that the study of common generalized graded objects, such
as Lie algebras, Grassmann algebras, superspaces, etc., although it has been
widely conducted\cite
{rittenberg78a,rittenberg78b,scheunert79,scheunert83a,scheunert83b,scheunert83c,lukierski78}%
, can be reduced to that of the corresponding superobjects (${\Bbb Z}_{2}$%
-graded), through a change of the commutation factor\cite{scheunert79,
marcinek91}. But this only means that any theorem that is true for
superobjects is also true for the generalized, arbitrarily graded objects,
which is a good thing. We claim that the commutation factor has physical
relevance in itself, and thus that the generalized objects can describe
objects whose properties are different from those of the analogous
superobjects. The commutation properties of operators describing particles
are at the origin of some of their most important features, that is, their
bosonic or fermionic statistics. The objects that we will describe here have
some properties of ordinary bosons and fermions, but with additional
features that could be useful in the modelization of quark fields.

\section{${\Bbb Z}_{n}^{3}$-graded Grassmann algebra}

Among the abelian groups that could be chosen to grade a generalized
Grassmann\label{pargrassmann} algebra, ${\Bbb Z}_{n}^{3}$ groups seem to be
very particular in being the only groups to induce a Grassmann algebra that
is maximally symmetric and includes fermionic, bosonic, and {\em other}
types of variables\cite{leroy97}. No abelian group composed of more than
three cyclic groups is able to produce a Grassmann algebra that puts them on
an equal footing, and any group composed of less than three cyclic groups
gives an ordinary Grassmann algebra.

We label the three grading groups with the letters $r$, $g$ and $b$ (for
red, green and blue) for reasons that will become clearer in the sequel. The
commutation factor for the generalized Grassmann algebra is then 
\[
\varepsilon
(x,y)=(-1)^{x_{r}y_{r}+x_{g}y_{g}+x_{b}y_{b}}q^{x_{r}y_{g}-y_{r}x_{g}+x_{g}y_{b}-y_{g}x_{b}+x_{b}y_{r}-y_{b}x_{r}} 
\]
where the degrees $x$ and $y$ are expressed by three integers representing
their components on the three groups: 
\[
x=(x_{r},x_{g},x_{b})\text{\quad and\quad }y=(y_{r},y_{g},y_{b}) 
\]
and $q$ is an $n^{th}$ root of unity.

The generalized Grassmann algebra is defined as a ${\Bbb Z}_{n}^{3}$-graded,
associative and $\varepsilon $-commutative algebra.

Its generators could be limited to elements of degree $1$ in only one of the
three colors, but we choose to include generators of degree $(\pm 1,\pm
1,\pm 1)$, which are fermionic (anticommuting) generators, as can be seen in
the commutation rules below, as well as generators of degree $-1$ in one of
the three colors. We summarize the degrees of the generators in the table
below: 
\[
\begin{tabular}{c|c|c|c|c|c|c|c|c|}
\cline{2-9}
& $\theta _{A_{r}}$ & $\theta _{A_{g}}$ & $\theta _{A_{b}}$ & $\bar{\theta}_{%
\bar{A}_{\bar{r}}}$ & $\bar{\theta}_{\bar{A}_{\bar{g}}}$ & $\bar{\theta}_{%
\bar{A}_{\bar{b}}}$ & $\eta _{a}$ & $\bar{\eta}_{\dot{a}}$ \\ \hline
\multicolumn{1}{|c|}{Rouge} & $1$ & $0$ & $0$ & $-1$ & $0$ & $0$ & $1$ & $-1$
\\ \hline
\multicolumn{1}{|c|}{Vert} & $0$ & $1$ & $0$ & $0$ & $-1$ & $0$ & $1$ & $-1$
\\ \hline
\multicolumn{1}{|c|}{Bleu} & $0$ & $0$ & $1$ & $0$ & $0$ & $-1$ & $1$ & $-1$
\\ \hline
\end{tabular}
\]

Some commutation rules, entirely defined by the commutation factor, are 
\[
\eta _{a}\eta _{b}=-\eta _{b}\eta _{a}\text{,\quad }\bar{\eta}_{\dot{a}}\bar{%
\eta}_{\dot{b}}=-\bar{\eta}_{\dot{b}}\bar{\eta}_{\dot{a}}\text{,\quad }\eta
_{a}\bar{\eta}_{\dot{b}}=-\bar{\eta}_{\dot{b}}\eta _{a}
\]
which are the commutation relations of ordinary fermionic variables; other
interesting commutation rules include: 
\[
\theta _{A_{r}}\theta _{A_{g}}=q\theta _{A_{g}}\theta _{A_{r}}\text{,\quad }%
\theta _{A_{g}}\theta _{A_{b}}=q\theta _{A_{b}}\theta _{A_{g}}\text{,\quad }%
\theta _{A_{b}}\theta _{A_{r}}=q\theta _{A_{r}}\theta _{A_{b}}
\]
\[
\bar{\theta}_{\bar{A}_{\bar{r}}}\bar{\theta}_{\bar{A}_{\bar{g}}}=q\bar{\theta%
}_{\bar{A}_{\bar{g}}}\bar{\theta}_{\bar{A}_{\bar{r}}}\text{,\quad }\bar{%
\theta}_{\bar{A}_{\bar{g}}}\bar{\theta}_{\bar{A}_{\bar{b}}}=q\bar{\theta}_{%
\bar{A}_{\bar{b}}}\bar{\theta}_{\bar{A}_{\bar{g}}}\text{,\quad }\bar{\theta}%
_{\bar{A}_{\bar{b}}}\bar{\theta}_{\bar{A}_{\bar{r}}}=q\bar{\theta}_{\bar{A}_{%
\bar{r}}}\bar{\theta}_{\bar{A}_{\bar{b}}}
\]
In a sector of a given color, the generators anticommute:
\[
\theta _{A_{r}}\theta _{B_{r}}=-\theta _{B_{r}}\theta _{A_{r}}
\]
Thus, the colored generators are nilpotent of rank two: $\theta
_{A_{r}}^{2}=\theta _{A_{g}}^{2}=\theta _{A_{b}}^{2}=\theta _{A_{\bar{r}%
}}^{2}=\theta _{A_{\bar{g}}}^{2}=\theta _{A_{\bar{b}}}^{2}=0$. Similarly,
colored generators commute with fermionic generators: 
\[
\theta _{A_{r}}\eta _{a}=-\eta _{a}\theta _{A_{r}}
\]

The most significant new feature of this algebra is that it includes a
purely fermionic and bosonic subalgebra, that is an ordinary Grassmann
subalgebra, that contains very particular combinations of the colored
generators, especially if $q$ is not a root of unity, that is, if the
grading group is ${\Bbb Z}^{3}$:

\begin{itemize}
\item  The fermionic (anticommuting) elements are:

\begin{itemize}
\item  The $\eta _{a}$ and $\bar{\eta}_{\dot{a}}$

\item  The products $\theta _{A_{r}}\theta _{A_{g}}\theta _{A_{b}}$ and $%
\bar{\theta}_{\bar{A}_{\bar{r}}}\bar{\theta}_{\bar{A}_{\bar{g}}}\bar{\theta}%
_{\bar{A}_{\bar{b}}}$ of three generators of {\em distinct colors}.

\item  The products of a bosonic element (see below) and a fermionic
element, or an odd number of the above.
\end{itemize}

\item  The bosonic elements (commuting with all other elements) are:

\begin{itemize}
\item  The products $\theta _{A_{r}}\bar{\theta}_{\bar{A}_{\bar{r}}}$, $%
\theta _{A_{g}}\bar{\theta}_{\bar{A}_{\bar{g}}}$ and $\theta _{A_{b}}\bar{%
\theta}_{\bar{A}_{\bar{b}}}$ of a generator and another of opposite color.

\item  The products of an even number of fermionic elements

\item  The products of bosonic elements.
\end{itemize}
\end{itemize}

This list is exhaustive, and one can note that the only possible
combinations of colored variables remind strongly of the observable
combinations of quarks in QCD. This property will have its analogue in the
irreducible representations of the Poincar\'{e} superalgebra.\ It is well
known that a reasonable field theory in a four dimensional minkowskian
space-time cannot feature observable particles whose creation and
annihilation operators do not obey ordinary commutation and anticommutation
rules\cite{spinstat}. Thus, in a field theory based on the objects that we
introduce here, the ``colored'' particles would be unobservable
individually, but the above combinations into fermions and bosons could be
observable. This could provide an algebraic model for the confinement of
quarks. We will develop this idea further in the next article.

\section{${\Bbb Z}_{n}^{3}$-graded superspace}

The ${\Bbb Z}_{n}^{3}$-graded superspaces are defined as a set $Q$ and an
associated family of bijections that are homogeneous in degree, the
coordinate maps, from $Q$ to the generalized Grassmann algebra\cite
{marcinek91}. In the case of ordinary ${\Bbb Z}_{2}$-graded superspaces,
there were only two degrees, and the question of the representation of all
degrees by the coordinates was irrelevant. Here, there can or can not be
coordinates of any possible degree.

One can define continuity, differentiation, analytic functions on these
superspaces, and the usual theorems generalize very well\cite{marcinek91}.
For example, 
\[
\frac{\partial }{\partial x_{i}}\frac{\partial }{\partial x_{j}}=\varepsilon
(d_{x_{i}},d_{x_{j}})\frac{\partial }{\partial x_{j}}\frac{\partial }{%
\partial x_{i}} 
\]
where $x_{i}$ and $x_{j}$ are two coordinates, and $d_{x_{i}}$ and $%
d_{x_{j}} $ are their degrees. The Leibniz rule generalizes as 
\[
\frac{\partial }{\partial x_{i}}(fg)=\frac{\partial f}{\partial x_{i}}%
g+\varepsilon (d_{x_{i}},d_{f})f\frac{\partial g}{\partial x_{i}} 
\]
if $f$ is a function of homogeneous degree $d_{f}$.

\section{${\Bbb Z}_{n}^{3}$-graded Poincar\'{e} superalgebras}

To generate a ${\Bbb Z}_{n}^{3}$-graded Poincar\'{e} superalgebra, we will
add to the ordinary Poincar\'{e} algebra a collection of generators of
arbitrary degrees to form an $\varepsilon $-Lie superalgebra\cite
{scheunert79}. The generalized supercommutator will be noted here as follows 
\[
\lbrack A,B]_{c}=AB-\varepsilon (d_{A},d_{B})BA
\]
We will focus on the smallest of these algebras that include the usual
minimal Poincar\'{e} superalgebra.

If we add to the Poincar\'{e} algebra the odd generators of a usual Poincar%
\'{e} superalgebra, we have to give them a suitable degree in our
generalized grading group. Obviously, the possible degrees are $\pm 1$ in
each color, like in the Grassmann algebra. The supercommutators of two
supertranslations must fall into the even part of the algebra, that is the
Poincar\'{e} algebra. Thus, the non-zero supercommutation relations of the
Poincar\'{e} superalgebra must be reflected here by colored supercommutation
relations between generators of {\em opposite} degrees (so that the result
is of degree $0$). Therefore, the two component and four component
formulations of Poincar\'{e} superalgebras won't generalize equivalently.

\subsection{Generalized two component formulation}

First, we'll generalize the two component formulation by adding $2$
generators $Q_{1},Q_{2}$ of degree $1$ in each color (that we'll call
``white'' generators) and $2$ generators $\bar{Q}_{\dot{1}},\bar{Q}_{\dot{2}%
} $ of degree $-1$ in each color (that we'll call ``antiwhite'' generators).
These generators will give the odd part of the ordinary Poincar\'{e}
superalgebra. Similarly, we'll add two generators and two ``antigenerators''
in each color. Finally, we will not assume anything \`{a} priori on the
commutation relations that don't give a zero degree result, but we'll try to
keep the algebra as small as possible. In summary, we have the following
generators: 
\[
\begin{tabular}{r|c|c|c|c|c|c|c|c|}
\cline{2-9}
& $Q_{1},Q_{2}$ & $\bar{Q}_{\dot{1}},\bar{Q}_{\dot{2}}$ & $%
Q_{1_{r}},Q_{2_{r}}$ & $\bar{Q}_{\dot{1}_{\bar{r}}},\bar{Q}_{\dot{2}_{\bar{r}%
}}$ & $Q_{1_{g}},Q_{2_{g}}$ & $\bar{Q}_{\dot{1}_{\bar{g}}},\bar{Q}_{\dot{2}_{%
\bar{g}}}$ & $Q_{1_{b}},Q_{2_{b}}$ & $\bar{Q}_{\dot{1}_{\bar{b}}},\bar{Q}_{%
\dot{2}_{\bar{b}}}$ \\ \hline
\multicolumn{1}{|r|}{red} & $1$ & $-1$ & $1$ & $-1$ & $0$ & $0$ & $0$ & $0$
\\ \hline
\multicolumn{1}{|r|}{green} & $1$ & $-1$ & $0$ & $0$ & $1$ & $-1$ & $0$ & $0$
\\ \hline
\multicolumn{1}{|r|}{blue} & $1$ & $-1$ & $0$ & $0$ & $0$ & $0$ & $1$ & $-1$
\\ \hline
\end{tabular}
\]

As in the case of Lie superalgebras, the elements of any sector of the
algebra of a given degree must form the basis for a representation of the
Poincar\'{e} algebra. We'll choose our generators here so that the
representation for positive (resp. negative) degree sectors are the left
(resp. right) handed irreducible two component representations of the
Lorentz algebra. The translations are trivially represented. In other words,
if $\alpha >\beta $, and if $d$ is any degree among $(1,1,1)$, $(-1,-1,-1)$, 
$r=(1,0,0)$, $\bar{r}=(-1,0,0)$, $g$, $\bar{g}$, $b$, $\bar{b}$, we have,
the ${\bf \sigma }$ being Pauli matrices:
\begin{eqnarray*}
\lbrack M_{\alpha \beta },Q_{i_{d}}]_{c} &=&-\frac{\hslash }{2i}%
\sum_{j_{d}=1}^{2}({\bf \sigma }_{\alpha }{\bf \sigma }_{\beta
})_{i_{d}j_{d}}Q_{j_{d}} \\
\lbrack M_{\alpha \beta },\bar{Q}_{i_{d}}]_{c} &=&-\frac{\hslash }{2i}%
\sum_{j_{d}=1}^{2}({\bf \sigma }_{\alpha }{\bf \sigma }_{\beta
})_{i_{d}j_{d}}^{*}\bar{Q}_{j_{d}}
\end{eqnarray*}
and if $\alpha <\beta $, 
\begin{eqnarray}
\lbrack M_{\alpha \beta },Q_{i_{d}}]_{c} &=&\frac{\hslash }{2i}%
\sum_{j_{d}=1}^{2}({\bf \sigma }_{\beta }{\bf \sigma }_{\alpha
})_{i_{d}j_{d}}Q_{j_{d}}  \label{comm_rot_suptrans} \\
\lbrack M_{\alpha \beta },\bar{Q}_{i_{d}}]_{c} &=&\frac{\hslash }{2i}%
\sum_{j_{d}=1}^{2}({\bf \sigma }_{\beta }{\bf \sigma }_{\alpha
})_{i_{d}j_{d}}^{*}\bar{Q}_{j_{d}}  \nonumber
\end{eqnarray}
And of course, 
\begin{equation}
\lbrack P_{\mu },Q_{i_{d}}]_{c}=[P_{\mu },\bar{Q}_{j_{d^{\prime }}}]_{c}=0
\label{comm_trans_suptrans}
\end{equation}

If we want the white and anti-white generators to behave like the
supertranslations, the commutator of two white ---or two antiwhite---
generators must be equal to zero. Similarly, we'll suppose that the
commutator of two generators of the same color is equal to zero, which will
keep the size of the algebra minimal.

Let us first compute the commutation relations of two generators of opposite
degrees. The result must be an element of the Poincar\'{e} algebra. The
generalized Jacobi identity\cite{scheunert79} and the commutation relations (%
\ref{comm_trans_suptrans}) give 
\[
\lbrack P_{\mu },[Q_{i_{d}},\bar{Q}_{j_{-d}}]_{c}]_{c}=0
\]
Thus, $[Q_{i_{d}},\bar{Q}_{j_{-d}}]_{c}$ must decompose along the
translations $P_{\mu }$. Another application of the Jacobi identity with the
rotations, and of the relations (\ref{comm_rot_suptrans}) give the
coefficients of this decomposition: 
\[
\lbrack Q_{i_{d}},\bar{Q}_{j_{-d}}]_{c}=\kappa _{d}\sum_{\mu =1}^{4}({\bf %
\sigma }_{\mu })_{i_{d}j_{-d}}P^{\mu }
\]
In supersymmetry, $\kappa _{d}$ is usually fixed to the value $2$, but for
the moment, we'll allow for different values of this parameter for each
degree $d\in \{(1,1,1),r,g,b\}$.

It is clear from these relations that the zero degree, the white and the
anti-white generated sectors form a subalgebra that is really a Poincar\'{e}
superalgebra (the colored commutator reduces in these sectors to the
supercommutator).

We still have to compute the commutation relations of two generators of
different and non-opposite degrees. We can reduce the dimension of these
bicolored sectors as low as $4$ while maintaining the Jacobi identities
true, by supposing that each of them is generated by four generators $%
R_{\alpha _{d}}$, where $d$ is any bicolor degree among $r+g$, $g+b$, $b+r$, 
$b+\bar{r}$, $r+\bar{g}$, $g+\bar{b}$ and their opposites, and that the
following commutation relations hold true: if $(d,d^{\prime })\in
\{(r,g),(g,b),(b,r),(\bar{r},\bar{g}),(\bar{g},\bar{b}),(\bar{b},\bar{r})\}$%
, 
\[
\lbrack Q_{a_{d}},Q_{b_{d^{\prime }}}]_{c}=\frac{1-q}{2}\sqrt{\kappa
_{d}\kappa _{d^{\prime }}}\sum_{a_{d+d^{\prime }}=1}^{4}({\bf \sigma }%
_{a_{d+d^{\prime }}}{\bf )}_{a_{d}b_{d^{\prime }}}R_{a_{d+d^{\prime }}} 
\]
If $(d,d^{\prime })\in \{(r,\bar{g}),(g,\bar{b}),(b,\bar{r}),(\bar{r},g),(%
\bar{g},b),(\bar{b},r)\}$, 
\[
\lbrack Q_{a_{d}},Q_{b_{d^{\prime }}}]_{c}=\frac{1-q^{-1}}{2}\sqrt{\kappa
_{d}\kappa _{d^{\prime }}}\sum_{a_{d+d^{\prime }}=1}^{4}({\bf \sigma }%
_{a_{d+d^{\prime }}})_{a_{d}b_{d^{\prime }}}R_{a_{d+d^{\prime }}} 
\]
We also have 
\begin{eqnarray*}
\lbrack Q_{a},Q_{b_{\bar{r}}}]_{c} &=&-\sqrt{\kappa _{1}\kappa _{r}}%
\sum_{a_{g+b}=1}^{4}({\bf \sigma }_{a_{g+b}})_{ab_{\bar{r}}}R_{a_{g+b}} \\
\lbrack Q_{a},Q_{b_{\bar{g}}}]_{c} &=&-\sqrt{\kappa _{1}\kappa _{g}}%
\sum_{a_{b+r}=1}^{4}({\bf \sigma }_{a_{b+r}})_{ab_{\bar{g}}}R_{a_{b+r}} \\
\lbrack Q_{a},Q_{b_{\bar{b}}}]_{c} &=&-\sqrt{\kappa _{1}\kappa _{b}}%
\sum_{a_{r+g}=1}^{4}({\bf \sigma }_{a_{r+g}})_{ab_{\bar{b}}}R_{a_{r+g}}
\end{eqnarray*}
and 
\begin{eqnarray*}
\lbrack Q_{\dot{a}},Q_{b_{r}}]_{c} &=&-\sqrt{\kappa _{1}\kappa _{r}}\sum_{a_{%
\bar{g}+\bar{b}}=1}^{4}({\bf \sigma }_{a_{\bar{g}+\bar{b}}})_{\dot{a}%
b_{r}}R_{a_{\bar{g}+\bar{b}}} \\
\lbrack Q_{\dot{a}},Q_{b_{g}}]_{c} &=&-\sqrt{\kappa _{1}\kappa _{g}}\sum_{a_{%
\bar{b}+\bar{r}}=1}^{4}({\bf \sigma }_{a_{\bar{b}+\bar{r}}})_{\dot{a}%
b_{g}}R_{a_{\bar{b}+\bar{r}}} \\
\lbrack Q_{\dot{a}},Q_{b_{b}}]_{c} &=&-\sqrt{\kappa _{1}\kappa _{b}}\sum_{a_{%
\bar{r}+\bar{g}}=1}^{4}({\bf \sigma }_{a_{\bar{r}+\bar{g}}})_{\dot{a}%
b_{b}}R_{a_{\bar{r}+\bar{g}}}
\end{eqnarray*}
Finally, for $\alpha <\beta $, the Jacobi identity gives 
\[
\lbrack M_{\alpha \beta },R_{a_{d+d^{\prime }}}]_{c}=\frac{\hslash }{i}(\eta
_{\alpha a_{d+d^{\prime }}}R_{\beta _{d+d^{\prime }}}-\eta _{\beta
a_{d+d^{\prime }}}R_{\alpha _{d+d^{\prime }}}) 
\]
The $d+d^{\prime }$ index clearly does not indicate the degree of $\alpha $
and $\beta $ indices, but that of $R$.

The only other commutation relations that are not equal to zero are the
usual 
\begin{eqnarray*}
\lbrack M_{\alpha \beta },M_{\eta \lambda }]_{c} &=&\frac{{\bf \hslash }}{i}%
(\eta _{\alpha \eta }M_{\beta \lambda }-\eta _{\alpha \lambda }M_{\beta \eta
}-\eta _{\beta \eta }M_{\alpha \lambda }+\eta _{\beta \lambda }M_{\alpha
\eta }) \\
\lbrack M_{\alpha \beta },P_{\mu }]_{c} &=&\frac{{\bf \hslash }}{i}(\eta
_{\alpha \mu }P_{\beta }-\eta _{\beta \mu }P_{\alpha })
\end{eqnarray*}

\subsection{Generalized four component formulation}

In the generalization of this formulation, we use four generators in each
color, anticolor, as well as in white and antiwhite. The notations are
basically the same as in the two component section, except that the indices
run from $1$ to $4$ instead of from $1$ to $2$. We choose these generators
so that the representation of the Poincar\'{e} algebra in the colored
sectors is a spinorial representation where the $M_{\alpha \beta }$ are
represented by 
\[
\frac{\hslash }{2i}\widetilde{({\bf \gamma }_{\alpha }{\bf \gamma }_{\beta })%
}
\]
where the tilda is the transposition operation and the ${\bf \gamma }%
_{\alpha }$ are Dirac matrices. The translations are trivially represented.

Like in the case of the two component formulation, we also introduce sets of
four generators $R_{\alpha _{d}}$ in the bicolor sectors. The Jacobi
identity, and the limitation to the minimal case where the algebra {\em is}
the vector space spanned by these generators (the $M_{\alpha \beta }$, $%
P_{\mu }$, $Q_{a_{d}}$ and $R_{\alpha _{d}}$) give us the following
commutation relations for $\alpha \neq \beta $: 
\begin{eqnarray*}
\lbrack M_{\alpha \beta },M_{\eta \lambda }]_{c} &=&\frac{{\bf \hslash }}{i}%
(\eta _{\alpha \eta }M_{\beta \lambda }-\eta _{\alpha \lambda }M_{\beta \eta
}-\eta _{\beta \eta }M_{\alpha \lambda }+\eta _{\beta \lambda }M_{\alpha
\eta }) \\
\lbrack M_{\alpha \beta },P_{\mu }]_{c} &=&\frac{{\bf \hslash }}{i}(\eta
_{\alpha \mu }P_{\beta }-\eta _{\beta \mu }P_{\alpha }) \\
\lbrack M_{\alpha \beta },Q_{i_{d}}]_{c} &=&\frac{\hslash }{2i}%
\sum_{b_{d}=1}^{4}({\bf \gamma }_{\alpha }{\bf \gamma }_{\beta
})_{i_{d}b_{d}}Q_{b_{d}} \\
\lbrack Q_{i_{d}},Q_{j_{-d}}]_{c} &=&-\kappa _{d}\sum_{\mu =1}^{4}({\bf %
\gamma }^{\mu }{\bf C})_{i_{d}j_{-d}}P_{\mu } \\
\lbrack Q_{a_{d}},Q_{b_{d^{\prime }}}]_{c} &=&\frac{1-q}{2}\sqrt{\kappa
_{d}\kappa _{d^{\prime }}}\sum_{a_{d+d^{\prime }}=1}^{4}({\bf \gamma }%
^{a_{d+d^{\prime }}}{\bf C})_{a_{d}b_{d^{\prime }}}R_{a_{d+d^{\prime }}} \\
\text{for }(d,d^{\prime }) &\in &\{(r,g),(g,b),(b,r),(\bar{r},\bar{g}),(\bar{%
g},\bar{b}),(\bar{b},\bar{r})\} \\
\lbrack Q_{a_{d}},Q_{b_{d^{\prime }}}]_{c} &=&\frac{1-q^{-1}}{2}\sqrt{\kappa
_{d}\kappa _{d^{\prime }}}\sum_{a_{d+d^{\prime }}=1}^{4}({\bf \gamma }%
^{a_{d+d^{\prime }}}{\bf C})_{a_{d}b_{d^{\prime }}}R_{a_{d+d^{\prime }}} \\
\text{for }(d,d^{\prime }) &\in &\{(r,\bar{g}),(g,\bar{b}),(b,\bar{r}),(\bar{%
r},g),(\bar{g},b),(\bar{b},r)\} \\
\lbrack Q_{a},Q_{b_{\bar{r}}}]_{c} &=&-\sqrt{\kappa _{1}\kappa _{r}}%
\sum_{a_{g+b}=1}^{4}({\bf \gamma }^{a_{g+b}}{\bf C})_{ab_{\bar{r}%
}}R_{a_{g+b}} \\
\lbrack Q_{a},Q_{b_{\bar{g}}}]_{c} &=&-\sqrt{\kappa _{1}\kappa _{g}}%
\sum_{a_{b+r}=1}^{4}({\bf \gamma }^{a_{b+r}}{\bf C})_{ab_{\bar{g}%
}}R_{a_{b+r}} \\
\lbrack Q_{a},Q_{b_{\bar{b}}}]_{c} &=&-\sqrt{\kappa _{1}\kappa _{b}}%
\sum_{a_{r+g}=1}^{4}({\bf \gamma }^{a_{r+g}}{\bf C})_{ab_{\bar{b}%
}}R_{a_{r+g}} \\
\lbrack Q_{\dot{a}},Q_{b_{r}}]_{c} &=&-\sqrt{\kappa _{1}\kappa _{r}}\sum_{a_{%
\bar{g}+\bar{b}}=1}^{4}({\bf \gamma }^{a_{\bar{g}+\bar{b}}}{\bf C})_{\dot{a}%
b_{r}}R_{a_{\bar{g}+\bar{b}}} \\
\lbrack Q_{\dot{a}},Q_{b_{g}}]_{c} &=&-\sqrt{\kappa _{1}\kappa _{g}}\sum_{a_{%
\bar{b}+\bar{r}}=1}^{4}({\bf \gamma }^{a_{\bar{b}+\bar{r}}}{\bf C})_{\dot{a}%
b_{g}}R_{a_{\bar{b}+\bar{r}}} \\
\lbrack Q_{\dot{a}},Q_{b_{b}}]_{c} &=&-\sqrt{\kappa _{1}\kappa _{b}}\sum_{a_{%
\bar{r}+\bar{g}}=1}^{4}({\bf \gamma }^{a_{\bar{r}+\bar{g}}}{\bf C})_{\dot{a}%
b_{b}}R_{a_{\bar{r}+\bar{g}}} \\
\lbrack M_{\alpha \beta },R_{a_{d+d^{\prime }}}]_{c} &=&\frac{\hslash }{i}%
(\eta _{\alpha a_{d+d^{\prime }}}R_{\beta _{d+d^{\prime }}}-\eta _{\beta
a_{d+d^{\prime }}}R_{\alpha _{d+d^{\prime }}})
\end{eqnarray*}

In these relations, ${\bf C}$ is the charge conjugation matrix.

\section{Representations of the ${\Bbb Z}_{n}^{3}$-graded Poincar\'{e}
superalgebras}

The above study of the minimal generalized Poincar\'{e} superalgebra has led
us to algebras where only $21$ degrees are present: $0$, $1$, $\bar{1}$, $r$%
, $g$, $b$, $\bar{r}$, $\bar{g}$, $\bar{b}$, $r+g$, $g+b$, $b+r$, $\bar{r}+%
\bar{g}$, $\bar{g}+\bar{b}$, $\bar{b}+\bar{r}$, $\bar{r}+g$, $\bar{g}+b$, $%
\bar{b}+r$, $r+\bar{g}$, $g+\bar{b}$, $b+\bar{r}$. We have to find a block
structure for the representation that reproduces the grading rules. In this
section, we consider only the representation of the four-component ${\Bbb Z}%
_{n}^{3}$-graded Poincar\'{e} superalgebra.

The diagonal blocks must represent transformations of degree $0$, that is
the Poincar\'{e} transformations. At least one of these block
representations must be faithful. On the other hand, the commutator of a
generator of any color and a generator of opposite color gives a linear
combination of translations. Thus, any block line or column corresponding to
a faithful representation of the Poincar\'{e} algebra must contain blocks of 
{\em all}{\bf \ }colors $1$, $r$, $g$, $b$ and their opposites. The smallest
structure meeting all requirements is $24\times 24$ by blocks, and its
actual size is $100\times 100$. This block structure is 
\[
\left( 
\begin{array}{cc}
{\bf A} & {\bf B} \\ 
{\bf C} & {\bf D}
\end{array}
\right) 
\]
where the structure of ${\bf A,B,C,D}$, the degrees associated with each
block, when this degree is expressed in the algebra, being: 
\[
{\bf A=}\left( 
\begin{array}{cccc}
0 & r+g & g+b & b+r \\ 
\bar{r}+\bar{g} & 0 & b+\bar{r} & \bar{g}+b \\ 
\bar{g}+\bar{b} & \bar{b}+r & 0 & r+\bar{g} \\ 
\bar{b}+\bar{r} & g+\bar{b} & \bar{r}+g & 0
\end{array}
\right) 
\]
where the blocks are $5\times 5$. 
\[
{\bf B=}\left( 
\begin{array}{cccccccccccccccccccc}
r & g & b & \bar{r} & \bar{g} & \bar{b} & 1 & \bar{1} &  &  &  &  &  &  &  & 
&  &  &  &  \\ 
\bar{g} & \bar{r} &  &  &  & \bar{1} & b &  & \bar{b} &  &  & g & r &  &  & 
&  & 1 &  &  \\ 
& \bar{b} & \bar{g} & \bar{1} &  &  & r &  &  & \bar{r} &  &  &  & b & g & 
&  &  & 1 &  \\ 
\bar{b} &  & \bar{r} &  & \bar{1} &  & g &  &  &  & \bar{g} &  &  &  &  & r
& b &  &  & 1
\end{array}
\right) 
\]
where each block is $5\times 4$%
\[
{\bf C=}\left( 
\begin{array}{cccc}
\bar{r} & g &  & b \\ 
\bar{g} & r & b &  \\ 
\bar{b} &  & g & r \\ 
r &  & 1 &  \\ 
g &  &  & 1 \\ 
b & 1 &  &  \\ 
\bar{1} & \bar{b} & \bar{r} & \bar{g} \\ 
1 &  &  &  \\ 
& b &  &  \\ 
&  & r &  \\ 
&  &  & g \\ 
& \bar{g} &  &  \\ 
& \bar{r} &  &  \\ 
&  & \bar{b} &  \\ 
&  & \bar{g} &  \\ 
&  &  & \bar{r} \\ 
&  &  & \bar{b} \\ 
& \bar{1} &  &  \\ 
&  & \bar{1} &  \\ 
&  &  & \bar{1}
\end{array}
\right) 
\]
where each block is $4\times 5$. ${\bf D}$ is a square matrix constituted of 
$20\times 20$ $4\times 4$ blocks that are all equal to zero, except for the
diagonal blocks, which are of degree zero. The blocks whose degrees have not
been represented in these block structures are always equal to zero.

A matrix and its block structure representing an element $a$ of the
generalized Poincar\'{e} algebra will be noted 
\[
{\bf \Gamma }(a)=({\bf \Gamma }_{i,j}(a))\Sb 0\leqslant i\leqslant 23  \\ %
0\leqslant j\leqslant 23  \endSb 
\]

The faithful representations of the Poincar\'{e} algebra will be in the four
diagonal blocks of ${\bf A}$: 
\[
{\bf \Gamma }_{0,0}(M_{\alpha \beta })={\bf \Gamma }_{1,1}(M_{\alpha \beta
})={\bf \Gamma }_{2,2}(M_{\alpha \beta })={\bf \Gamma }_{3,3}(M_{\alpha
\beta })=\left( 
\begin{array}{cc}
{\bf M}_{\alpha \beta } & {\bf 0} \\ 
{\bf 0} & 0
\end{array}
\right) 
\]
where the ${\bf M}_{\alpha \beta }$ matrices are defined by 
\begin{equation}
({\bf M}_{\alpha \beta })_{\mu \nu }=\frac{\hslash }{i}(\delta _{\beta \mu
}\eta _{\alpha \nu }-\delta _{\alpha \mu }\eta _{\beta \nu })
\label{reprmalphabeta}
\end{equation}
where $\alpha $, $\beta $, $\lambda $, $\mu =1,\ldots ,4$. For the
translations, 
\[
{\bf \Gamma }_{0,0}(P_{\mu })={\bf \Gamma }_{1,1}(P_{\mu })={\bf \Gamma }%
_{2,2}(P_{\mu })={\bf \Gamma }_{3,3}(P_{\mu })=\left( 
\begin{array}{cc}
{\bf 0} & -\frac{i\hslash }{\lambda }{\bf \delta }_{\mu } \\ 
{\bf 0} & {\bf 0}
\end{array}
\right) \equiv {\bf P}_{\mu } 
\]
where the ${\bf \delta }_{\mu }$ are the four $4\times 1$ matrices defined
by 
\[
({\bf \delta }_{\mu })_{\alpha 1}=\delta _{\alpha \mu } 
\]
and $\lambda $ is a real constant with the dimensions of a length.

For the other representations of the Poincar\'{e} algebra, we will choose a
spinorial representation: for $i>3$, 
\begin{eqnarray*}
{\bf \Gamma }_{i,i}(M_{\alpha \beta }) &=&-\frac{\hslash }{2i}{\bf \gamma }%
_{\alpha }{\bf \gamma }_{\beta } \\
{\bf \Gamma }_{i,i}(P_{\mu }) &=&{\bf 0}
\end{eqnarray*}

The supertranslations $Q_{a}$ are represented by the matrices with the
following non-zero blocks: 
\begin{eqnarray*}
{\bf \Gamma }_{0,10}(Q_{a}) &=&{\bf \Gamma }_{1,21}(Q_{a})={\bf \Gamma }%
_{2,22}(Q_{a})={\bf \Gamma }_{3,23}(Q_{a})=\sqrt{\kappa _{1}}{\bf B}_{a} \\
{\bf \Gamma }_{11,0}(Q_{a}) &=&{\bf \Gamma }_{9,1}(Q_{a})={\bf \Gamma }%
_{7,2}(Q_{a})={\bf \Gamma }_{8,3}(Q_{a})=\sqrt{\kappa _{1}}{\bf C}_{a}
\end{eqnarray*}
where ${\bf B}_{a}$ and ${\bf C}_{a}$ are the matrices defined by 
\begin{eqnarray*}
{\bf B}_{a} &=&\left( 
\begin{array}{c}
-\left( \frac{\hslash }{\lambda }\right) ^{1/2}e^{i\pi /4}{\bf U}_{a} \\ 
{\bf 0}
\end{array}
\right) \\
{\bf C}_{a} &=&\left( 
\begin{array}{cc}
{\bf 0} & -\left( \frac{\hslash }{\lambda }\right) ^{1/2}e^{i\pi /4}{\bf %
\delta }_{a}
\end{array}
\right)
\end{eqnarray*}
The ${\bf 0}$ in ${\bf B}_{a}$ is a $1\times 4$ zero block and the ${\bf 0}$
in ${\bf C}_{a}$ is a $4\times 4$ zero block. The ${\bf U}_{a}$ are four $%
4\times 4$ matrices defined by 
\[
({\bf U}_{a})_{\alpha b}=({\bf \gamma }^{\alpha }{\bf C)}_{ab} 
\]

Similarly, the supertranslations $Q_{\dot{a}}$ are represented by the $%
4\times 5$ blocks 
\begin{eqnarray*}
{\bf \Gamma }_{0,11}(Q_{\dot{a}}) &=&{\bf \Gamma }_{1,9}(Q_{\dot{a}})={\bf %
\Gamma }_{2,7}(Q_{\dot{a}})={\bf \Gamma }_{3,8}(Q_{\dot{a}})=\sqrt{\kappa
_{1}}{\bf B}_{\dot{a}} \\
{\bf \Gamma }_{10,0}(Q_{\dot{a}}) &=&{\bf \Gamma }_{21,1}(Q_{\dot{a}})={\bf %
\Gamma }_{22,2}(Q_{\dot{a}})={\bf \Gamma }_{23,3}(Q_{\dot{a}})=\sqrt{\kappa
_{1}}{\bf C}_{\dot{a}}
\end{eqnarray*}
and the colored supertranslations are represented by the blocks 
\begin{eqnarray*}
{\bf \Gamma }_{0,4}(Q_{i_{r}}) &=&{\bf \Gamma }_{1,16}(Q_{i_{r}})={\bf %
\Gamma }_{2,10}(Q_{i_{r}})={\bf \Gamma }_{3,19}(Q_{i_{r}})=\sqrt{\kappa _{r}}%
{\bf B}_{i_{r}} \\
{\bf \Gamma }_{7,0}(Q_{i_{r}}) &=&{\bf \Gamma }_{5,1}(Q_{i_{r}})={\bf \Gamma 
}_{13,2}(Q_{i_{r}})={\bf \Gamma }_{6,3}(Q_{i_{r}})=\sqrt{\kappa _{r}}{\bf C}%
_{i_{r}} \\
{\bf \Gamma }_{0,5}(Q_{i_{g}}) &=&{\bf \Gamma }_{1,15}(Q_{i_{g}})={\bf %
\Gamma }_{2,18}(Q_{i_{g}})={\bf \Gamma }_{3,10}(Q_{i_{g}})=\sqrt{\kappa _{g}}%
{\bf B}_{i_{g}} \\
{\bf \Gamma }_{8,0}(Q_{i_{g}}) &=&{\bf \Gamma }_{4,1}(Q_{i_{g}})={\bf \Gamma 
}_{6,2}(Q_{i_{g}})={\bf \Gamma }_{14,3}(Q_{i_{g}})=\sqrt{\kappa _{g}}{\bf C}%
_{i_{g}} \\
{\bf \Gamma }_{0,6}(Q_{i_{b}}) &=&{\bf \Gamma }_{1,10}(Q_{i_{b}})={\bf %
\Gamma }_{2,17}(Q_{i_{b}})={\bf \Gamma }_{3,20}(Q_{i_{b}})=\sqrt{\kappa _{b}}%
{\bf B}_{i_{b}} \\
{\bf \Gamma }_{9,0}(Q_{i_{b}}) &=&{\bf \Gamma }_{12,1}(Q_{i_{b}})={\bf %
\Gamma }_{5,2}(Q_{i_{b}})={\bf \Gamma }_{4,3}(Q_{i_{b}})=\sqrt{\kappa _{b}}%
{\bf C}_{i_{b}}
\end{eqnarray*}
and 
\begin{eqnarray*}
{\bf \Gamma }_{0,7}(Q_{i_{\bar{r}}}) &=&{\bf \Gamma }_{1,5}(Q_{i_{\bar{r}}})=%
{\bf \Gamma }_{2,13}(Q_{i_{\bar{r}}})={\bf \Gamma }_{3,6}(Q_{i_{\bar{r}}})=%
\sqrt{\kappa _{\bar{r}}}{\bf B}_{i_{\bar{r}}} \\
{\bf \Gamma }_{4,0}(Q_{i_{\bar{r}}}) &=&{\bf \Gamma }_{16,1}(Q_{i_{\bar{r}%
}})={\bf \Gamma }_{10,2}(Q_{i_{\bar{r}}})={\bf \Gamma }_{19,3}(Q_{i_{\bar{r}%
}})=\sqrt{\kappa _{\bar{r}}}{\bf C}_{i_{\bar{r}}} \\
{\bf \Gamma }_{0,8}(Q_{i_{\bar{g}}}) &=&{\bf \Gamma }_{1,4}(Q_{i_{\bar{g}}})=%
{\bf \Gamma }_{2,6}(Q_{i_{\bar{g}}})={\bf \Gamma }_{3,14}(Q_{i_{\bar{g}}})=%
\sqrt{\kappa _{\bar{g}}}{\bf B}_{i_{\bar{g}}} \\
{\bf \Gamma }_{5,0}(Q_{i_{\bar{g}}}) &=&{\bf \Gamma }_{15,1}(Q_{i_{\bar{g}%
}})={\bf \Gamma }_{18,2}(Q_{i_{\bar{g}}})={\bf \Gamma }_{10,3}(Q_{i_{\bar{g}%
}})=\sqrt{\kappa _{\bar{g}}}{\bf C}_{i_{\bar{g}}} \\
{\bf \Gamma }_{0,9}(Q_{i_{\bar{b}}}) &=&{\bf \Gamma }_{1,12}(Q_{i_{\bar{b}%
}})={\bf \Gamma }_{2,5}(Q_{i_{\bar{b}}})={\bf \Gamma }_{3,4}(Q_{i_{\bar{b}%
}})=\sqrt{\kappa _{\bar{b}}}{\bf B}_{i_{\bar{b}}} \\
{\bf \Gamma }_{6,0}(Q_{i_{\bar{b}}}) &=&{\bf \Gamma }_{10,1}(Q_{i_{\bar{b}%
}})={\bf \Gamma }_{17,2}(Q_{i_{\bar{b}}})={\bf \Gamma }_{20,3}(Q_{i_{\bar{b}%
}})=\sqrt{\kappa _{\bar{b}}}{\bf C}_{i_{\bar{b}}}
\end{eqnarray*}

Finally, the $R_{a_{d+d^{\prime }}}$ are represented by the matrices whose
non zero blocks are: 
\begin{eqnarray*}
{\bf \Gamma }_{0,1}(R_{a_{r+g}}) &=&{\bf P}_{a_{r+g}}\text{;\quad }{\bf %
\Gamma }_{0,2}(R_{a_{g+b}})={\bf P}_{a_{g+b}}\text{;\quad }{\bf \Gamma }%
_{0,3}(R_{a_{b+r}})={\bf P}_{a_{b+r}} \\
{\bf \Gamma }_{1,0}(R_{a_{\bar{r}+\bar{g}}}) &=&{\bf P}_{a_{\bar{r}+\bar{g}}}%
\text{;\quad }{\bf \Gamma }_{2,0}(R_{a_{\bar{g}+\bar{b}}})={\bf P}_{a_{\bar{g%
}+\bar{b}}}\text{;\quad }{\bf \Gamma }_{3,0}(R_{a_{\bar{b}+\bar{r}}})={\bf P}%
_{a_{\bar{b}+\bar{r}}} \\
{\bf \Gamma }_{1,2}(R_{a_{b+\bar{r}}}) &=&{\bf P}_{a_{b+\bar{r}}}\text{%
;\quad }{\bf \Gamma }_{1,3}(R_{a_{\bar{g}+b}})={\bf P}_{a_{\bar{g}+b}}\text{%
;\quad }{\bf \Gamma }_{2,3}(R_{a_{r+\bar{g}}})={\bf P}_{a_{r+\bar{g}}} \\
{\bf \Gamma }_{2,1}(R_{a_{\bar{b}+r}}) &=&{\bf P}_{a_{\bar{b}+r}}\text{%
;\quad }{\bf \Gamma }_{3,1}(R_{a_{g+\bar{b}}})={\bf P}_{a_{g+\bar{b}}}\text{%
;\quad }{\bf \Gamma }_{3,2}(R_{a_{\bar{r}+g}})={\bf P}_{a_{\bar{r}+g}}
\end{eqnarray*}

The matrix ${\bf \Gamma }(a)$ representing an arbitrary element $a$ of the
generalized Poincar\'{e} superalgebra can then be written 
\begin{eqnarray*}
{\bf \Gamma }(a) &=&\frac{i}{\hslash }\left[ \sum_{1\leqslant \alpha <\beta
\leqslant 4}\omega ^{\alpha \beta }{\bf \Gamma }(M_{\alpha \beta
})+\sum_{\mu =1}^{4}t^{\mu }{\bf \Gamma }(P_{\mu })+\sum\Sb (d+d^{\prime
})\in T \\ a_{d+d^{\prime }}=1 \endSb ^{4}u^{a_{d+d^{\prime }}}{\bf \Gamma }%
(R_{a_{d+d^{\prime }}})\right] + \\
&&+\hslash ^{-1/2}e^{-i\pi /4}\sum\Sb d\in \{1,\bar{1},r,g,b,\bar{r},\bar{g},%
\bar{b}\} \\ a_{d}=1 \endSb ^{4}\psi ^{a_{d}}{\bf \Gamma }(Q_{a_{d}})
\end{eqnarray*}
where 
\[
T=\{r+g,g+b,b+r,\bar{r}+\bar{g},\bar{g}+\bar{b},\bar{b}+\bar{r},r+\bar{g},g+%
\bar{b},b+\bar{r},\bar{r}+g,\bar{g}+b,\bar{b}+r\}
\]
$\omega ^{\alpha \beta }$ are six real dimensionless parameters, $t^{\mu }$
and $u^{a_{d+d^{\prime }}}$ are fifty-two parameters with the dimensions of
a length, and $\psi ^{a_{d}}$ are thirty-two parameters with the dimension
of the square root of a length, that are real in the case of the Majorana
representation of the Dirac matrices.

\section{${\Bbb Z}_{n}^{3}$-graded Poincar\'{e} supergroup}

It is possible to rewrite the representation of an arbitrary element of the
generalized Poincar\'{e} algebra as a supermatrix of degree $0$, introducing
Grassmann valued parameters: 
\begin{eqnarray}
{\bf \Gamma }(a) &=&\frac{i}{\hslash }\left[ \sum_{1\leqslant \alpha <\beta
\leqslant 4}\Omega ^{\alpha \beta }{\bf \Gamma }(M_{\alpha \beta
})+\sum_{\mu =1}^{4}T^{\mu }{\bf \Gamma }(P_{\mu })+\sum\Sb (d+d^{\prime
})\in T \\ a_{d+d^{\prime }}=1 \endSb ^{4}U^{a_{d+d^{\prime }}\#}{\bf \Gamma 
}(R_{a_{d+d^{\prime }}})\right]   \nonumber \\
&&+\frac{i}{\hslash ^{1/2}}\sum\Sb d\in \{1,\bar{1},r,g,b,\bar{r},\bar{g},%
\bar{b}\} \\ a_{d},b_{d}=1 \endSb ^{4}\zeta ^{a_{d}\#}({\bf \gamma }%
_{4})_{a_{d}b_{d}}{\bf \Gamma }(Q_{b_{d}})  \label{reppoinalg}
\end{eqnarray}
where $\Omega ^{\alpha \beta }$ are $6$ dimensionless parameters of degree $0
$, $T^{\mu }$ are $4$ parameters with the dimensions of a length and of
degree $0$, $U^{a_{d+d^{\prime }}}$ are $48$ parameters with the dimensions
of a length and of degrees $d+d^{\prime }$, $\zeta ^{a_{d}}$ are $32$
parameters with the dimensions of the square root of a length and of degrees 
$d$. The $\#$ operator is the adjoint operator of the generalized Grassmann
algebra, which is defined by 
\begin{eqnarray*}
(x.{\bf 1})^{\#} &=&x^{*}.{\bf 1} \\
(\eta _{a})^{\#} &=&-i\eta _{a}\text{,\quad }(\bar{\eta}_{\dot{a}})^{\#}=-i%
\bar{\eta}_{\dot{a}} \\
(\theta _{A_{d}})^{\#} &=&iq\theta _{A_{d}}\text{,\quad }(\bar{\theta}_{%
\bar{A}_{\bar{d}}})^{\#}=iq\bar{\theta}_{\bar{A}_{\bar{d}}} \\
(XY)^{\#} &=&Y^{\#}X^{\#}
\end{eqnarray*}

The choice of parameters (\ref{reppoinalg}) ensures that they transform in
the same way as the operators they multiply under a transformation of the
Dirac matrices.

A representation of the generalized Poincar\'{e} supergroup is obtained by
exponentiating these matrices.

The generalized superspace parametrizing the generalized Poincar\'{e}
supergroup has the following dimensions 
\[
\begin{array}{r}
(\dim _{d};d\in \{\text{degrees}\})=(10_{0},4_{1},4_{\bar{1}%
},4_{r},4_{g},4_{b},4_{\bar{r}},4_{\bar{g}},4_{\bar{b}}, \\ 
4_{r+g},4_{g+b},4_{b+r},4_{\bar{r}+\bar{g}},4_{\bar{g}+\bar{b}},4_{\bar{b}+%
\bar{r}}, \\ 
4_{\bar{r}+g},4_{\bar{g}+b},4_{\bar{b}+r},4_{r+\bar{g}},4_{g+\bar{b}},4_{b+%
\bar{r}})
\end{array}
\]

An element of the Poincar\'{e} supergroup specified by the parameters $%
\Omega ^{\alpha \beta }$, $T^{\mu }$, $\zeta ^{a_{d}}$, and $%
U^{a_{d+d^{\prime }}}$ will be noted $[{\bf \Lambda }({\bf \Omega })\mid 
{\bf T}\mid {\bf \zeta }\mid {\bf U}]$, and its representation ${\bf \Gamma }%
([{\bf \Lambda }({\bf \Omega })\mid {\bf T}\mid {\bf \zeta }\mid {\bf U}])$.

For an element whose only nonvanishing coordinates are the $\Omega ^{\alpha
\beta }$, 
\[
{\bf \Gamma }([{\bf \Lambda }({\bf \Omega })\mid {\bf 0}\mid {\bf 0}\mid 
{\bf 0}])=\left( 
\begin{array}{cc}
{\bf A} & {\bf 0} \\ 
{\bf 0} & {\bf B}
\end{array}
\right) 
\]
where 
\begin{eqnarray*}
{\bf A} &=&\left( 
\begin{array}{cccccccc}
{\bf \Lambda }({\bf \Omega }) & {\bf 0} & {\bf 0} & {\bf 0} & {\bf 0} & {\bf %
0} & {\bf 0} & {\bf 0} \\ 
{\bf 0} & 1 & {\bf 0} & {\bf 0} & {\bf 0} & {\bf 0} & {\bf 0} & {\bf 0} \\ 
{\bf 0} & {\bf 0} & {\bf \Lambda }({\bf \Omega }) & {\bf 0} & {\bf 0} & {\bf %
0} & {\bf 0} & {\bf 0} \\ 
{\bf 0} & {\bf 0} & {\bf 0} & 1 & {\bf 0} & {\bf 0} & {\bf 0} & {\bf 0} \\ 
{\bf 0} & {\bf 0} & {\bf 0} & {\bf 0} & {\bf \Lambda }({\bf \Omega }) & {\bf %
0} & {\bf 0} & {\bf 0} \\ 
{\bf 0} & {\bf 0} & {\bf 0} & {\bf 0} & {\bf 0} & 1 & {\bf 0} & {\bf 0} \\ 
{\bf 0} & {\bf 0} & {\bf 0} & {\bf 0} & {\bf 0} & {\bf 0} & {\bf \Lambda }(%
{\bf \Omega }) & {\bf 0} \\ 
{\bf 0} & {\bf 0} & {\bf 0} & {\bf 0} & {\bf 0} & {\bf 0} & {\bf 0} & 1
\end{array}
\right) \\
{\bf B} &=&\left( 
\begin{array}{cccc}
{\bf \Gamma }^{\limfunc{spin}}({\bf \Lambda }({\bf \Omega })) & {\bf 0} & 
\cdots & {\bf 0} \\ 
{\bf 0} & {\bf \Gamma }^{\limfunc{spin}}({\bf \Lambda }({\bf \Omega })) & 
\ddots & \vdots \\ 
\vdots & \ddots & \ddots & {\bf 0} \\ 
{\bf 0} & \cdots & {\bf 0} & {\bf \Gamma }^{\limfunc{spin}}({\bf \Lambda }(%
{\bf \Omega }))
\end{array}
\right)
\end{eqnarray*}
where the ${\bf \Lambda }({\bf \Omega })$ are the Lorentz supermatrices
obtained from the Lorentz matrices by replacing the real parameters $\omega
^{\alpha \beta }$ with their zero degree grassmannian counterparts, the $%
\Omega ^{\alpha \beta }$; the ${\bf \Gamma }^{\limfunc{spin}}({\bf \Lambda }%
(\Omega ))$ are spinorial representations of the Lorentz supermatrices.

The square of all elements ${\bf M}$ of the algebra whose $\Omega $
coordinates are equal to zero vanish. Moreover, two matrices representing
elements of the same degree (excluding zero) have a vanishing product. Thus $%
\exp ({\bf M})={\bf 1}+{\bf M}$ and if we choose to note 
\[
{\cal T}=\left( 
\begin{array}{cc}
{\bf 1}_{4} & \sum\limits_{\mu =1}^{4}T^{\mu }\frac{{\bf \delta }_{\mu }}{%
\lambda } \\ 
{\bf 0} & 1
\end{array}
\right) 
\]
we have 
\begin{eqnarray*}
{\bf \Gamma }([{\bf 1} &\mid &{\bf T}\mid {\bf 0}\mid {\bf 0}])={\bf 1}%
_{100}+\frac{i}{\hslash }\sum_{\mu =1}^{4}T^{\mu }{\bf \Gamma }(P_{\mu }) \\
&=&\left( 
\begin{array}{ccccc}
{\cal T} & {\bf 0} & \cdots & \cdots & {\bf 0} \\ 
{\bf 0} & {\cal T} & \ddots &  & \vdots \\ 
\vdots & \ddots & {\cal T} & \ddots & \vdots \\ 
\vdots &  & \ddots & {\cal T} & {\bf 0} \\ 
{\bf 0} & \cdots & \cdots & {\bf 0} & {\bf 1}_{80}
\end{array}
\right) \\
{\bf \Gamma }([{\bf 1} &\mid &{\bf 0}\mid {\bf \zeta }^{d}\mid {\bf 0}])=%
{\bf 1}_{100}+\frac{i}{\hslash ^{1/2}}\sum_{a_{d},b_{d}=1}^{4}\zeta
^{a_{d}\#}({\bf \gamma }_{4})_{a_{d}b_{d}}{\bf \Gamma }(Q_{b_{d}}) \\
{\bf \Gamma }([{\bf 1} &\mid &{\bf 0}\mid {\bf 0}\mid {\bf U}])={\bf 1}%
_{100}+\frac{i}{\hslash }\sum\Sb (d+d^{\prime })\in T  \\ a_{d+d^{\prime
}}=1  \endSb ^{4}U^{a_{d+d^{\prime }}\#}{\bf \Gamma }(R_{a_{d+d^{\prime }}})
\end{eqnarray*}

An arbitrary element $[\Lambda \mid {\bf T}\mid {\bf \zeta }\mid {\bf U}]$
of the Poincar\'{e} supergroup can be written as 
\[
\begin{array}{r}
\lbrack {\bf \Lambda }\mid {\bf T}\mid {\bf \zeta }\mid {\bf U}]=[{\bf 1}%
\mid {\bf 0}\mid {\bf 0}\mid {\bf U}]\prod\limits_{d\in \{1,r,g,b,\bar{1},%
\bar{r},\bar{g},\bar{b}\}}[{\bf 1}\mid {\bf 0}\mid {\bf \zeta }^{d}\mid {\bf %
0}]\times  \\ 
\times [{\bf 1}\mid {\bf T}\mid {\bf 0}\mid {\bf 0}][{\bf \Lambda }\mid {\bf %
0}\mid {\bf 0}\mid {\bf 0}]
\end{array}
\]
Another order convention for the product would amount to a phase change in
the parameters.

The product of two elements $[{\bf \Lambda }\mid {\bf T}\mid {\bf \zeta }%
\mid {\bf U}]$ and $[{\bf \Lambda }^{\prime }\mid {\bf T}^{\prime }\mid {\bf %
\zeta }^{\prime }\mid {\bf U}^{\prime }]$ of the supergroup is defined by 
\[
{\bf \Gamma }([{\bf \Lambda }\mid {\bf T}\mid {\bf \zeta }\mid {\bf U}][{\bf %
\Lambda }^{\prime }\mid {\bf T}^{\prime }\mid {\bf \zeta }^{\prime }\mid 
{\bf U}^{\prime }])={\bf \Gamma }([{\bf \Lambda }\mid {\bf T}\mid {\bf \zeta 
}\mid {\bf U}]){\bf \Gamma }([{\bf \Lambda }^{\prime }\mid {\bf T}^{\prime
}\mid {\bf \zeta }^{\prime }\mid {\bf U}^{\prime }])
\]
which gives 
\begin{equation}
\begin{array}{l}
\lbrack {\bf \Lambda }\mid {\bf T}\mid {\bf \zeta }\mid {\bf U}][{\bf %
\Lambda }^{\prime }\mid {\bf T}^{\prime }\mid {\bf \zeta }^{\prime }\mid 
{\bf U}^{\prime }]= \\ 
=[{\bf \Lambda \Lambda }^{\prime }\mid {\bf T+\Lambda T}^{\prime }+{\bf \tau 
}\mid {\bf \zeta +\Gamma }^{\limfunc{spin}}({\bf \Lambda }){\bf \zeta }%
^{\prime }\mid {\bf U+\Lambda U^{\prime }+\rho }]
\end{array}
\label{mulgrpoinca}
\end{equation}
where ${\bf T}^{\prime }$ is the vector whose components are the $T^{\prime
\mu }$. ${\bf \Gamma }^{\limfunc{spin}}({\bf \Lambda }){\bf \zeta }^{\prime }
$ stands for the set of all parameters ${\bf \Gamma }^{\limfunc{spin}}({\bf %
\Lambda }){\bf \zeta }^{\prime d}$ for each degree $d$ in $\{1,r,g,b,\bar{1},%
\bar{r},\bar{g},\bar{b}\}$, where ${\bf \zeta }^{\prime d}$ is the vector
whose components are the $\zeta ^{\prime a_{d}}$. The same notations are
used for ${\bf \Lambda U^{\prime }}$. ${\bf \tau }$ is defined by 
\[
\tau ^{\mu }=\sum_{d\in \{1,r,g,b,\bar{1},\bar{r},\bar{g},\bar{b}\}}i(%
\widetilde{{\bf \zeta }^{d}})^{\#}{\bf \gamma }_{4}{\bf \gamma }^{\mu }{\bf %
\Gamma }^{\limfunc{spin}}({\bf \Lambda }){\bf \zeta }^{\prime -d}
\]
and ${\bf \rho }$ by 
\begin{eqnarray*}
\rho ^{a_{r+g}} &=&i(\widetilde{{\bf \zeta }^{r}})^{\#}{\bf \gamma }_{4}{\bf %
\gamma }^{a_{r+g}}{\bf \Gamma }^{\limfunc{spin}}({\bf \Lambda }){\bf \zeta }%
^{\prime g}+i(\widetilde{{\bf \zeta }^{g}})^{\#}{\bf \gamma }_{4}{\bf \gamma 
}^{a_{r+g}}{\bf \Gamma }^{\limfunc{spin}}({\bf \Lambda }){\bf \zeta }%
^{\prime r}+ \\
&&+i(\widetilde{{\bf \zeta }^{\bar{1}}})^{\#}{\bf \gamma }_{4}{\bf \gamma }%
^{a_{r+g}}{\bf \Gamma }^{\limfunc{spin}}({\bf \Lambda }){\bf \zeta }^{\prime
b}+i(\widetilde{{\bf \zeta }^{b}})^{\#}{\bf \gamma }_{4}{\bf \gamma }%
^{a_{r+g}}{\bf \Gamma }^{\limfunc{spin}}({\bf \Lambda }){\bf \zeta }^{\prime 
\bar{1}} \\
\rho ^{a_{\bar{r}+\bar{g}}} &=&i(\widetilde{{\bf \zeta }^{\bar{r}}})^{\#}%
{\bf \gamma }_{4}{\bf \gamma }^{a_{\bar{r}+\bar{g}}}{\bf \Gamma }^{\limfunc{%
spin}}({\bf \Lambda }){\bf \zeta }^{\prime \bar{g}}+i(\widetilde{{\bf \zeta }%
^{\bar{g}}})^{\#}{\bf \gamma }_{4}{\bf \gamma }^{a_{\bar{r}+\bar{g}}}{\bf %
\Gamma }^{\limfunc{spin}}({\bf \Lambda }){\bf \zeta }^{\prime \bar{r}}+ \\
&&+i(\widetilde{{\bf \zeta }^{1}})^{\#}{\bf \gamma }_{4}{\bf \gamma }^{a_{%
\bar{r}+\bar{g}}}{\bf \Gamma }^{\limfunc{spin}}({\bf \Lambda }){\bf \zeta }%
^{\prime \bar{b}}+i(\widetilde{{\bf \zeta }^{\bar{b}}})^{\#}{\bf \gamma }_{4}%
{\bf \gamma }^{a_{\bar{r}+\bar{g}}}{\bf \Gamma }^{\limfunc{spin}}({\bf %
\Lambda }){\bf \zeta }^{\prime 1} \\
\rho ^{a_{r+\bar{g}}} &=&i(\widetilde{{\bf \zeta }^{r}})^{\#}{\bf \gamma }%
_{4}{\bf \gamma }^{a_{r+\bar{g}}}{\bf \Gamma }^{\limfunc{spin}}({\bf \Lambda 
}){\bf \zeta }^{\prime \bar{g}}+i(\widetilde{{\bf \zeta }^{\bar{g}}})^{\#}%
{\bf \gamma }_{4}{\bf \gamma }^{a_{r+\bar{g}}}{\bf \Gamma }^{\limfunc{spin}}(%
{\bf \Lambda }){\bf \zeta }^{\prime r} \\
\rho ^{a_{\bar{r}+g}} &=&i(\widetilde{{\bf \zeta }^{\bar{r}}})^{\#}{\bf %
\gamma }_{4}{\bf \gamma }^{a_{\bar{r}+g}}{\bf \Gamma }^{\limfunc{spin}}({\bf %
\Lambda }){\bf \zeta }^{\prime g}+i(\widetilde{{\bf \zeta }^{g}})^{\#}{\bf %
\gamma }_{4}{\bf \gamma }^{a_{\bar{r}+g}}{\bf \Gamma }^{\limfunc{spin}}({\bf %
\Lambda }){\bf \zeta }^{\prime \bar{r}}
\end{eqnarray*}
and the four equivalent formulas in other colors.

An immediate consequence is that 
\[
\lbrack {\bf \Lambda }\mid {\bf T}\mid {\bf \zeta }\mid {\bf U}]^{-1}=[{\bf %
\Lambda }^{-1}\mid -{\bf \Lambda }^{-1}{\bf T}\mid {\bf -\Gamma }^{\limfunc{%
spin}}({\bf \Lambda }^{-1}){\bf \zeta }\mid {\bf -\Lambda }^{-1}{\bf %
U^{\prime }}] 
\]

\section{Action of the ${\Bbb Z}_{n}^{3}$-graded Poincar\'{e} supergroup on
the ${\Bbb Z}_{n}^{3}$-graded superspace}

The multiplication rule of two elements of the Poincar\'{e} supergroup
admits as a particular case 
\[
\lbrack {\bf 1}\mid {\bf X}\mid {\bf \Xi }\mid {\bf \Omega }][{\bf \Lambda }%
\mid {\bf 0}\mid {\bf 0}\mid {\bf 0}]=[{\bf \Lambda }\mid {\bf X}\mid {\bf %
\Xi }\mid {\bf \Omega }] 
\]
Thus, all elements of a left orthochronous coset of the Poincar\'{e}
supergroup with homogeneous orthochronous Lorentz transformations formed
from a given element $[{\bf 1}\mid {\bf X}\mid {\bf \Xi }\mid {\bf \Omega }]$
of the Poincar\'{e} supergroup have the {\em same} translational parts
specified by ${\bf X}$, ${\bf \Xi }$ et ${\bf \Omega }$. The coset is thus
entirely defined by ${\bf X}$, ${\bf \Xi }$ and ${\bf \Omega }$. The action
of an arbitrary element $[{\bf \Lambda }\mid {\bf T}\mid {\bf \zeta }\mid 
{\bf U}]$ of the Poincar\'{e} supergroup on the representant $[{\bf 1}\mid 
{\bf X}\mid {\bf \Xi }\mid {\bf \Omega }]$ of the coset is given by 
\[
\lbrack {\bf \Lambda }\mid {\bf T}\mid {\bf \zeta }\mid {\bf U}][{\bf 1}\mid 
{\bf X}\mid {\bf \Xi }\mid {\bf \Omega }]=[{\bf \Lambda }\mid {\bf \Lambda
X+T}+{\bf \tau }\mid {\bf \Gamma }^{\limfunc{spin}}({\bf \Lambda }){\bf \Xi
+\zeta }\mid {\bf \Lambda \Omega +U+\rho }] 
\]
where 
\[
\tau ^{\mu }=\sum_{d\in \{1,r,g,b,\bar{1},\bar{r},\bar{g},\bar{b}\}}i(%
\widetilde{{\bf \zeta }^{d}})^{\#}{\bf \gamma }_{4}{\bf \gamma }^{\mu }{\bf %
\Gamma }^{\limfunc{spin}}({\bf \Lambda }){\bf \Xi }^{-d} 
\]
and 
\begin{eqnarray*}
\rho ^{a_{r+g}} &=&i(\widetilde{{\bf \zeta }^{r}})^{\#}{\bf \gamma }_{4}{\bf %
\gamma }^{a_{r+g}}{\bf \Gamma }^{\limfunc{spin}}({\bf \Lambda }){\bf \Xi }%
^{g}+i(\widetilde{{\bf \zeta }^{g}})^{\#}{\bf \gamma }_{4}{\bf \gamma }%
^{a_{r+g}}{\bf \Gamma }^{\limfunc{spin}}({\bf \Lambda }){\bf \Xi }^{r}+ \\
&&+i(\widetilde{{\bf \zeta }^{\bar{1}}})^{\#}{\bf \gamma }_{4}{\bf \gamma }%
^{a_{r+g}}{\bf \Gamma }^{\limfunc{spin}}({\bf \Lambda }){\bf \Xi }^{b}+i(%
\widetilde{{\bf \zeta }^{b}})^{\#}{\bf \gamma }_{4}{\bf \gamma }^{a_{r+g}}%
{\bf \Gamma }^{\limfunc{spin}}({\bf \Lambda }){\bf \Xi }^{\bar{1}} \\
\rho ^{a_{\bar{r}+\bar{g}}} &=&i(\widetilde{{\bf \zeta }^{\bar{r}}})^{\#}%
{\bf \gamma }_{4}{\bf \gamma }^{a_{\bar{r}+\bar{g}}}{\bf \Gamma }^{\limfunc{%
spin}}({\bf \Lambda }){\bf \Xi }^{\bar{g}}+i(\widetilde{{\bf \zeta }^{\bar{g}%
}})^{\#}{\bf \gamma }_{4}{\bf \gamma }^{a_{\bar{r}+\bar{g}}}{\bf \Gamma }^{%
\limfunc{spin}}({\bf \Lambda }){\bf \Xi }^{\bar{r}}+ \\
&&+i(\widetilde{{\bf \zeta }^{1}})^{\#}{\bf \gamma }_{4}{\bf \gamma }^{a_{%
\bar{r}+\bar{g}}}{\bf \Gamma }^{\limfunc{spin}}({\bf \Lambda }){\bf \Xi }^{%
\bar{b}}+i(\widetilde{{\bf \zeta }^{\bar{b}}})^{\#}{\bf \gamma }_{4}{\bf %
\gamma }^{a_{\bar{r}+\bar{g}}}{\bf \Gamma }^{\limfunc{spin}}({\bf \Lambda })%
{\bf \Xi }^{1} \\
\rho ^{a_{r+\bar{g}}} &=&i(\widetilde{{\bf \zeta }^{r}})^{\#}{\bf \gamma }%
_{4}{\bf \gamma }^{a_{r+\bar{g}}}{\bf \Gamma }^{\limfunc{spin}}({\bf \Lambda 
}){\bf \Xi }^{\bar{g}}+i(\widetilde{{\bf \zeta }^{\bar{g}}})^{\#}{\bf \gamma 
}_{4}{\bf \gamma }^{a_{r+\bar{g}}}{\bf \Gamma }^{\limfunc{spin}}({\bf %
\Lambda }){\bf \Xi }^{r} \\
\rho ^{a_{\bar{r}+g}} &=&i(\widetilde{{\bf \zeta }^{\bar{r}}})^{\#}{\bf %
\gamma }_{4}{\bf \gamma }^{a_{\bar{r}+g}}{\bf \Gamma }^{\limfunc{spin}}({\bf %
\Lambda }){\bf \Xi }^{g}+i(\widetilde{{\bf \zeta }^{g}})^{\#}{\bf \gamma }%
_{4}{\bf \gamma }^{a_{\bar{r}+g}}{\bf \Gamma }^{\limfunc{spin}}({\bf \Lambda 
}){\bf \Xi }^{\bar{r}}
\end{eqnarray*}
In other words, the action of the transformation $[{\bf \Lambda }\mid {\bf T}%
\mid {\bf \zeta }\mid {\bf U}]$ on the coset defined by ${\bf X}$, ${\bf \Xi 
}$ and ${\bf \Omega }$ results in the coset defined by ${\bf \Lambda X+T}+%
{\bf \tau }$, ${\bf \Gamma }^{\limfunc{spin}}({\bf \Lambda }){\bf \Xi +\zeta 
}$ and ${\bf \Lambda \Omega +U+\rho }$.

Thus, we define the action of the generalized Poincar\'{e} supergroup on a
point of the superspace defined by the coordinates ${\bf X}$, ${\bf \Xi }$
and ${\bf \Omega }$ as its action on the coset defined by the same
coordinates. The dimensions of the superspace are given by: 
\begin{eqnarray*}
D &=&(4_{0},4_{1},4_{\bar{1}},4_{r},4_{g},4_{b},4_{\bar{r}},4_{\bar{g}},4_{%
\bar{b}}, \\
&&4_{r+g},4_{g+b},4_{b+r},4_{\bar{r}+\bar{g}},4_{\bar{g}+\bar{b}},4_{\bar{b}+%
\bar{r}},4_{\bar{r}+g},4_{\bar{g}+b},4_{\bar{b}+r},4_{r+\bar{g}},4_{g+\bar{b}%
},4_{b+\bar{r}})
\end{eqnarray*}

An element $[{\bf \Lambda }\mid {\bf T}\mid {\bf \zeta }\mid {\bf U}]$ of
the Poincar\'{e} supergroup transforms the point $({\bf X},{\bf \Xi ,\Omega }%
)$ into the point $({\bf X}^{\prime },{\bf \Xi }^{\prime }{\bf ,\Omega }%
^{\prime })$, where 
\begin{eqnarray*}
{\bf X}^{\prime } &=&{\bf \Lambda X+T}+{\bf \tau } \\
{\bf \Xi }^{\prime } &=&{\bf \Gamma }^{\limfunc{spin}}({\bf \Lambda }){\bf %
\Xi +\zeta } \\
{\bf \Omega }^{\prime } &=&{\bf \Lambda \Omega +U+\rho }
\end{eqnarray*}
We'll also note $({\bf X}^{\prime },{\bf \Xi }^{\prime }{\bf ,\Omega }%
^{\prime })=[{\bf \Lambda }\mid {\bf T}\mid {\bf \zeta }\mid {\bf U}]({\bf X}%
,{\bf \Xi ,\Omega )}$.

Of course, the consecutive action of two transformations on a point is
equivalent to the action of the product of the transformations.

This rule includes the following particular cases. If we apply a homogeneous
Lorentz transformation $[{\bf \Lambda }\mid {\bf 0}\mid {\bf 0}\mid {\bf 0}]$%
, 
\begin{eqnarray*}
{\bf X}^{\prime } &=&{\bf \Lambda X} \\
{\bf \Xi }^{\prime } &=&{\bf \Gamma }^{\limfunc{spin}}({\bf \Lambda }){\bf %
\Xi } \\
{\bf \Omega }^{\prime } &=&{\bf \Lambda \Omega }
\end{eqnarray*}
If we apply a translation $[{\bf 1}\mid {\bf T}\mid {\bf 0}\mid {\bf 0}]$, 
\begin{eqnarray*}
{\bf X}^{\prime } &=&{\bf X+T} \\
{\bf \Xi }^{\prime } &=&{\bf \Xi } \\
{\bf \Omega }^{\prime } &=&{\bf \Omega }
\end{eqnarray*}
If we apply a colored supertranslation $[{\bf 1}\mid {\bf 0}\mid {\bf \zeta }%
\mid {\bf 0}]$, 
\begin{eqnarray*}
{\bf X}^{\prime \mu } &=&{\bf X}^{\mu }+\sum_{d\in \{1,r,g,b,\bar{1},\bar{r},%
\bar{g},\bar{b}\}}i(\widetilde{{\bf \zeta }^{d}})^{\#}{\bf \gamma }_{4}{\bf %
\gamma }^{\mu }{\bf \Gamma }^{\limfunc{spin}}({\bf \Lambda }){\bf \Xi }^{-d}
\\
{\bf \Xi }^{\prime } &=&{\bf \Xi +\zeta } \\
{\bf \Omega }^{\prime } &=&{\bf \Omega +\rho }
\end{eqnarray*}
Finally, if we apply $[{\bf 1}\mid {\bf 0}\mid {\bf 0}\mid {\bf U}]$, 
\begin{eqnarray*}
{\bf X}^{\prime } &=&{\bf X} \\
{\bf \Xi }^{\prime } &=&{\bf \Xi } \\
{\bf \Omega }^{\prime } &=&{\bf \Omega +U}
\end{eqnarray*}

We'll call scalar superfield an analytic operator-valued function ${\bf \Phi 
}_{s}({\bf X},{\bf \Xi ,\Omega )}$ on the superspace.

The transformation operators $P([{\bf \Lambda }\mid {\bf T}\mid {\bf \zeta }%
\mid {\bf U}])$ for the scalar superfields are defined by the prescription 
\[
P([{\bf \Lambda }\mid {\bf T}\mid {\bf \zeta }\mid {\bf U}]){\bf \Phi }_{s}(%
{\bf X},{\bf \Xi ,\Omega )}P([{\bf \Lambda }\mid {\bf T}\mid {\bf \zeta }%
\mid {\bf U}])^{-1}={\bf \Phi }_{s}{\bf (}[{\bf \Lambda }\mid {\bf T}\mid 
{\bf \zeta }\mid {\bf U}]({\bf X},{\bf \Xi ,\Omega ))} 
\]

In the case of supertranslations, we'll use the notation 
\[
\delta _{{\bf \zeta }}{\bf \Phi }_{s}({\bf X},{\bf \Xi ,\Omega )}=\left[
P\left( \frac{i}{\hslash ^{1/2}}\sum\Sb d\in \{1,\bar{1},r,g,b,\bar{r},\bar{g%
},\bar{b}\}  \\ a_{d},b_{d}=1  \endSb ^{4}\zeta ^{a_{d}\#}({\bf \gamma }%
_{4})_{a_{d}b_{d}}{\bf \Gamma }(Q_{b_{d}})\right) ,{\bf \Phi }_{s}({\bf X},%
{\bf \Xi ,\Omega )}\right] 
\]

From the action of the supergroup on the superspace, we get 
\[
\begin{array}{l}
\lbrack P(Q_{a_{d}}),{\bf \Phi }_{s}({\bf X},{\bf \Xi ,\Omega )]=} \\ 
=\hslash ^{-1/2}\left\{ \dsum\limits\Sb d\in \{1,\bar{1},r,g,b,\bar{r},\bar{g%
},\bar{b}\}  \\ \mu =1  \endSb ^{4}({\bf \gamma }^{\mu }{\bf \Xi }%
^{d})_{a_{d}}\dfrac{\partial }{\partial X^{\mu }}+i\dsum\limits\Sb d\in \{1,%
\bar{1},r,g,b,\bar{r},\bar{g},\bar{b}\}  \\ b_{d}=1  \endSb %
^{4}C_{a_{d}b_{d}}\dfrac{\partial }{\partial \Xi ^{a_{d}}}\right\} {\bf \Phi 
}_{s}({\bf X},{\bf \Xi ,\Omega )} \\ 
\lbrack P(R_{a_{d+d^{\prime }}}),{\bf \Phi }_{s}({\bf X},{\bf \Xi ,\Omega )]=%
}\dfrac{\hslash }{i}\dfrac{\partial {\bf \Phi }_{s}({\bf X},{\bf \Xi ,\Omega
)}}{\partial \Omega ^{a_{d+d^{\prime }}}} \\ 
\lbrack P(M_{\alpha \beta }),{\bf \Phi }_{s}({\bf X},{\bf \Xi ,\Omega )]=}
\\ 
=\dfrac{\hslash }{i}\left\{ \left( X_{\alpha }\dfrac{\partial }{\partial
X^{\beta }}-X_{\beta }\dfrac{\partial }{\partial X^{\alpha }}\right) -\dfrac{%
1}{2}\dsum\limits\Sb d\in \{1,\bar{1},r,g,b,\bar{r},\bar{g},\bar{b}\}  \\ %
a_{d},b_{d}=1  \endSb ^{4}\Xi ^{a_{d}}({\bf \gamma }_{\alpha }{\bf \gamma }%
_{\beta })_{b_{d}a_{d}}\dfrac{\partial }{\partial \Xi ^{b_{d}}}\right\} {\bf %
\Phi }_{s}({\bf X},{\bf \Xi ,\Omega )} \\ 
\lbrack P(P_{\mu }),{\bf \Phi }_{s}({\bf X},{\bf \Xi ,\Omega )]=}\dfrac{%
\hslash }{i}\dfrac{\partial {\bf \Phi }_{s}({\bf X},{\bf \Xi ,\Omega )}}{%
\partial X^{\mu }}
\end{array}
\]

\section{Conclusion}

We have constructed here a generalized Poincar\'{e} superalgebra and the
corresponding supergroup based on the larger grading group ${\Bbb Z}_{n}^{3}$%
, as well as its action on the corresponding superspace. Even though these
constructions can be brought back to ordinary superstructures through a
change of the commutation factor, some properties appear clearly only with
the original commutation factor, which has some relevance in itself. This
will be shown in more details in the next article, where we will describe
the particle contents of the theory (especially their spin and statistics)
through the study of the irreducible representations of the Poincar\'{e}
superalgebra that has been developed here.

\end{document}